\documentclass[11pt]{article}
\usepackage[preprint]{acl}
\usepackage{times}
\usepackage{latexsym}
\usepackage[T1]{fontenc}

\usepackage[utf8]{inputenc}

\usepackage{microtype}

\usepackage{inconsolata}

\usepackage{graphicx}
\usepackage[most]{tcolorbox}
\tcbuselibrary{listings,breakable}

\title{Guiding LLM Peer Reviewers: The Impact of Score Anchors on Review Evidence and Accuracy}

\author{
Judita Preiss \and Yunhan Yang \\
School of Information, Journalism and Communication \\
The University of Sheffield \\
Sheffield, United Kingdom \\
\texttt{yunhan.yang@sheffield.ac.uk}
}

\begin{document}
\maketitle
\begin{abstract}
Large language models (LLMs) are increasingly used for research quality evaluation, with prior work exploring their scoring accuracy and the plausibility of review rationales. However, less is known about whether external score guidance changes the evidence presented in the generated review as well as the final score. This study uses 98 Allied Health Professions research outputs submitted for internal REF-style assessment, with specialist human review reports and adjudicated 1--4 reference scores. No-guidance baseline reviews are compared with oracle-guided reviews, where the supplied score is set to the rounded human reference score; extracted evaluation points are used to compare human and LLM evidence use. Using this design, oracle guidance improves scoring accuracy, with score-following checks showing that models do not simply copy the supplied score. Corrected score mismatches are associated with changes in the generated review frame, showing that the score signal can steer review rationales. This effect is direction-dependent: LLM reviews cover human strength or upgrade points more reliably than human weakness or downgrade points, with the weakest alignment for expert downgrade evidence. The results show that score-guided review generation can be evaluated at the level of review evidence, as well as the final score.
\end{abstract}

\section{Introduction}
\label{sec:introduction}
Large language models (LLMs)\footnote{Throughout this paper, the evaluated LLMs are generative decoder-only models used to produce review reports and scores. Annotation models are described separately in Section~\ref{sec:implementation}.} are increasingly being tested for scholarly evaluation tasks, including paper reviewing, review-quality assessment, and research quality prediction~\citep{zhou-etal-2024-llm-reviewer,liang2024can,thakkar2026large,wu-etal-2026-ai-peer-review}. In post-publication research quality evaluation, recent Research Excellence Framework (REF)-style studies have tested ChatGPT and other LLMs across different inputs, prompts, model settings, and scoring strategies, finding that LLM scores can correlate with expert or REF-based quality proxies~\citep{thelwall2024chatgpt-quality,thelwall2025settings-inputs,thelwall-yaghi2025fields,thelwall-jiang-bath2025medical,thelwall-yang2025probabilities,thelwall2025smaller,thelwall-mohammadi2026small-reasoning}. Such correlations suggest that LLMs may capture some ranking signal, but they do not show that models can assign accurate scores on the target evaluation scale. This motivates score-guided evaluation, where an accurate, external, score signal is used to guide the generative review. The key question is whether such guidance changes only the final score, or also the evidence used in the accompanying rationale.

This paper addresses this question using a REF-style post-publication evaluation, where already published outputs are assessed against criteria such as originality, significance, and rigour~\citep{ref2021results}. This setting is useful because the score and the written rationale are tightly linked: a review should not only assign a plausible quality level, but also justify that level using relevant strengths and limitations.

The empirical setting is a small but unusually informative internal REF-style dataset\footnote{University ethical approval is in place: details including application number will be provided in the final version.}: 98 anonymised outputs from UK Unit of Assessment 3, Allied Health Professions, with private expert review reports and overall scores from two to four specialist reviewers. The paper-level mean score was rounded to form the 1--4 human reference score. Score guidance was tested in two ways: a full guide-score sweep over all score options measures direct score-following, and an oracle-guidance condition supplied the human reference score for comparison with the no-guidance baseline.

This paper makes three contributions. First, it provides an evaluation protocol for distinguishing direct score copying from substantive changes in generated review rationales. Second, it provides a point-level framework for comparing human and LLM evidence use as score-raising or score-lowering support. Third, it identifies the review-frame shifts most associated with successful score correction: underestimates are more often corrected when guidance increases strength framing and reduces weakness framing, whereas overestimates are more often corrected when guidance reduces novelty/world-leading framing.

\section{Methods}
\label{sec:methods}
LLM outputs are analysed at two levels: the assigned score and the generated review content.

Score-level analysis used the two guidance settings introduced above. The guide-score sweep measured direct score-following by comparing supplied and final LLM scores. The oracle-guidance condition compared no-guidance and oracle-guided scores using exact-match accuracy and mean absolute error (MAE) against the human reference score. Scores below, equal to, or above the human reference score were labelled as underestimates, correct matches, or overestimates respectively. A score ``mismatch'' therefore means a mismatch with the human reference score, not an objectively incorrect judgement. 

For review-content analysis, no-guidance baseline outputs were first screened for shared error patterns to avoid analysing many model-hyperparameter combinations following \citet{kohli2026nine}. Three distinct configurations were retained. Within these configurations, each no-guidance review was paired with the corresponding oracle-guided review for the same paper and generation setting. This review pair was the main unit of text-level analysis. Pairs were grouped by score transition: corrected underestimate, persistent underestimate, corrected overestimate, or persistent overestimate.

The review text is annotated in two ways: (1) {\bf sentence-level review-frame labels} were assigned with zero-shot classification to estimate broad evaluative framing; (2) {\bf point-level analysis} treated each evaluation point as a discrete claim or judgement that supports, lowers, or qualifies a score. An LLM-as-a-judge approach extracts these points and labelled their theme, stance, and score implication, corresponding to evidence type, evaluative direction, and score direction respectively~\citep{preiss2025hybrid}. Human-point coverage recorded whether each human-review point was covered, partially covered, or not covered in the corresponding LLM review, encoded as 1, 0.5, and 0.

\section{Experiments}
\subsection{Settings}
The experiments used the 98 anonymised outputs described in Section~\ref{sec:introduction}. Three paper-input variants were prepared for each paper: title and abstract, title with introduction and conclusion, and full text. PDF anonymisation, text extraction, and section extraction details are provided in Appendix~\ref{appendix:dataset}.

Reviews were generated with Llama-3.1-8B, Falcon3-10B, and Llama-3.1-405B. Llama-3.1-8B and Falcon3-10B used three prompt variants (v4, v7, and v9); Llama-3.1-405B used v4 only because of generation cost. The prompt variants kept the same REF-style criteria and output fields, but changed the order of the task rules, paper text, and output-format instructions (see Appendix~\ref{appendix:prompt_templates}).

The full generation pool combined three paper-input variants, prompt variants, 11 hyperparameter settings, and guide scores from 1 to 4; the full hyperparameter list is reported in Appendix~\ref{appendix:screening}. Configuration screening was applied separately to each model and prompt setting using no-guidance baseline outputs, as described in Section~\ref{sec:methods}. The primary text-level analysis used Llama-3.1-8B prompt v4 with full-text input and three generation settings: temperature/top-p = 0.1/0.9, 0.3/0.8, and 0.5/0.9. Other model and prompt settings are used as robustness checks.

\subsection{Implementation details}\label{sec:implementation}
All LLM generations used a maximum generation length of 1024 new tokens. Sentence-level review-frame labels were produced with \texttt{MoritzLaurer/deberta-v3-large-zeroshot-v2.0}, a DeBERTa-v3-large NLI-based zero-shot classifier. The full zero-shot label set and hypothesis templates are provided in Appendix~\ref{appendix:text_features}.

Point-level analysis used Qwen3-32B as an LLM judge for evaluation-point extraction and cross-text coverage judgements. Qwen was selected because it comes from a different model family to Llama and Falcon, with different training data, tokenisation, and instruction tuning, making it a more independent annotation model. The point-level label set was developed after manually inspecting a subset of human review reports (see Appendix~\ref{appendix:text_features}).

LLM generation, Qwen-based annotation, and zero-shot classification were run on NVIDIA H100 GPUs. 

\subsection{Results}
\subsubsection{Oracle guidance changed scores without simple copying}
The first analysis tested whether oracle score guidance changed LLM scores. As shown in Table~\ref{tab:score_guidance_accuracy}, guidance improved exact-match accuracy across all selected full-text configurations, with the largest gains for Llama-405B v4, Llama-8B v7, and Falcon-10B v7. MAE also decreased in every setting, indicating that guided scores were closer to the human reference scores even when they were not exact matches.

\begin{table}[t]
\centering
\footnotesize
\setlength{\tabcolsep}{2.5pt}
\begin{tabular}{llrrrrrr}
\hline
Model & Pr. & Acc. & Acc. & $\Delta$ & MAE & MAE & $\Delta$ \\
      &     & base & guid. & Acc.     & base & guid. & MAE \\
\hline
Llama-8B & v4 & 0.490 & 0.551 & +0.061 & 0.544 & 0.456 & -0.088 \\
Llama-8B & v7 & 0.571 & 0.772 & +0.201 & 0.456 & 0.231 & -0.224 \\
Llama-8B & v9 & 0.452 & 0.554 & +0.102 & 0.592 & 0.483 & -0.109 \\
Falcon-10B & v4 & 0.531 & 0.721 & +0.190 & 0.473 & 0.282 & -0.190 \\
Falcon-10B & v7 & 0.473 & 0.667 & +0.194 & 0.537 & 0.333 & -0.204 \\
Falcon-10B & v9 & 0.622 & 0.796 & +0.173 & 0.378 & 0.204 & -0.173 \\
Llama-405B & v4 & 0.639 & 0.844 & +0.204 & 0.374 & 0.160 & -0.214 \\
\hline
\end{tabular}
\caption{Effect of oracle score guidance on score accuracy and error for the selected full-text settings.}
\label{tab:score_guidance_accuracy}
\end{table}

The score gains were not due to a direct copy of the supplied score: in a score-following check using guide scores from 1 to 4, the final LLM score exactly matched the supplied guide score in only 31.89\% of cases. The correlation between supplied and final scores was 0.32, and no paper-configuration pair copied the supplied score across all four guide-score values (1, 2, 3, and 4). Thus, the score signal acted as an anchor rather than a deterministic output constraint.

Score improvement was uneven across baseline mismatches (see Table~\ref{tab:error_correction}). Guidance corrected 31 of 68 baseline underestimates and 38 of 82 baseline overestimates, while 35 underestimates and 37 overestimates remain mismatched in the same direction. This motivates the text-level analysis below, which asks whether corrected and persistent mismatches differ in how the generated review content changes after guidance.

\begin{table}[t]
\centering
\small
\setlength{\tabcolsep}{3pt}
\begin{tabular}{lrrr}
\hline
Base error & $n$ pairs & Corrected & Persistent \\
\hline
Underestimate & 68 & 31 & 35 \\
Overestimate & 82 & 38 & 37 \\
\hline
\end{tabular}
\caption{Correction of baseline score mismatches after oracle guidance in the main text-analysis setting. Underestimate and overestimate are defined relative to the rounded human reference score.}
\label{tab:error_correction}
\end{table}
 
\subsubsection{Human and LLM reviews used different evidence frames}
Differences in the type of evidence used in human vs LLM baseline reviews can be seen in Table~\ref{tab:evidence_frames}, which groups Qwen-extracted evaluation points by theme and score implication. Human higher-score points were spread across novelty/originality, practical/clinical/policy/industrial impact, international/global significance, and methodology/rigour. LLM higher-score points showed a similar emphasis on novelty/originality, methodology/rigour, and practical/clinical/policy/industrial impact, but placed less emphasis on international/global significance.

\begin{table}[t]
\centering
\small
\setlength{\tabcolsep}{1pt}
\begin{tabular}{lrrrr}
\hline
Theme & Human$\uparrow$ & LLM$\uparrow$ & Human$\downarrow$ & LLM$\downarrow$ \\
\hline
Novelty/orig. & 23.7 & 29.3 & 4.3 & 7.2 \\
Practical impact & 23.7 & 24.2 & 14.5 & 3.6 \\
Intl./global significance & 22.8 & 11.6 & 12.9 & 2.0 \\
Method/rigour & 21.8 & 27.2 & 14.0 & 20.3 \\
Gen./context limitation & 0.7 & 0.7 & 21.5 & 27.5 \\
Incr./uncertain contribution & 0.5 & 0.4 & 14.5 & 12.0 \\
Evidence scale & 4.1 & 4.7 & 14.5 & 26.2 \\
Clarity/presentation & 1.84 & 1.39 & 3.23 & 0.68 \\
Other & 0.92 & 0.39 & 0.54 & 0.45 \\
\hline
\end{tabular}
\caption{Themes of score-raising and score-lowering evaluation points in human and LLM baseline reviews. Arrows indicate whether Qwen labelled each point as supporting a higher or lower score. Values are percentages within each source and score-implication group.}
\label{tab:evidence_frames}
\end{table}

The lower-score points show a different pattern. Human downgrade evidence included practical-impact limits and international-significance limits more often than LLM downgrade evidence. LLM lower-score points were instead more concentrated in generalisability limitation, methodology/rigour, and evidence scale. This point-level comparison suggests that human and LLM reviews did not only differ in topic coverage, but also in how similar themes were used to raise or constrain a score.

\subsubsection{Guidance steered review frames unevenly}
Changes in the LLM review frame after guidance were explored next. Table~\ref{tab:frame_change} uses sentence-level stance and review-frame labels, rather then the extracted evaluation points used in Table~\ref{tab:evidence_frames}. Corrected underestimates moved further toward a positive frame: strength framing increased more, weakness framing decreased more, and the strength--weakness balance shifted more clearly upward. Persistent underestimates showed weaker movement in the same direction.

\begin{table}[t]
\centering
\small
\setlength{\tabcolsep}{3pt}
\begin{tabular}{llrr}
\hline
Base dir. & Feature & Persist. & Corrected \\
\hline
Under. & Base weakness ratio & 0.167 & 0.105 \\
Under. & $\Delta$ strength--weakness & 0.022 & 0.125 \\
Under. & $\Delta$ strength ratio & 0.021 & 0.056 \\
Under. & $\Delta$ weakness ratio & -0.004 & -0.033 \\
Over. & Base strength ratio & 0.789 & 0.757 \\
Over. & $\Delta$ novelty/world-leading & 0.043 & -0.064 \\
Over. & $\Delta$ method weakness & -0.001 & 0.022 \\
\hline
\end{tabular}
\caption{Review-frame shifts in corrected and persistent score mismatches in the main text-analysis setting. Base dir. indicates whether the baseline LLM score was below or above the rounded human reference score. Values are medians.}
\label{tab:frame_change}
\end{table}

For overestimates, correction was associated with a clearer downgrading of the original positive frame. Corrected overestimates reduced novelty/world-leading framing and slightly increased methodological-weakness framing, while persistent overestimates retained or slightly increased novelty/world-leading language. Successful correction therefore usually involved a review-frame shift in the direction of the guided score.

Human evaluation points were then used as a more specific reference for evidence coverage. These points are specific claims extracted from human reports, rather than the three broad REF criteria themselves; the point-theme label set is described in Appendix~\ref{appendix:text_features}. Table~\ref{tab:coverage_patterns} summarises the strongest cross-setting coverage patterns. LLM reviews covered human upgrade/strength points relatively well, including in persistent underestimates. Coverage of human downgrade/weakness points in overestimated cases was much lower. Impact-related and methodology/rigour downgrade points were weakly covered in both corrected and persistent overestimates, suggesting a broader mismatch in how expert downgrade evidence was represented. For underestimates, the clearest gap was evidence-scale upgrade evidence, which was much less covered in persistent underestimates than in corrected underestimates.

Robustness checks across the additional selected model and prompt settings supported the same broad interpretation: corrected mismatches were usually associated with larger review-frame shifts, and human upgrade points were covered more consistently than downgrade points. Full setting-specific results are reported in Appendix~\ref{app:supp_results}.

\begin{table}[t]
\centering
\scriptsize
\setlength{\tabcolsep}{3pt}
\begin{tabular}{llrr}
\hline
Human point type & Compared cases & Corrected & Persistent \\
\hline
Upgrade/strength overall & Under. & 0.761 & 0.787 \\
Downgrade/weakness overall & Over. & 0.240 & 0.204 \\
Impact downgrade & Over. & 0.105 & 0.132 \\
Method/rigour downgrade & Over. & 0.102 & 0.160 \\
Evidence-scale upgrade & Under. & 0.697 & 0.336 \\
\hline
\end{tabular}
\caption{Cross-setting coverage of selected human evaluation-point types in corrected and persistent mismatches. Values are mean Qwen coverage scores.}
\label{tab:coverage_patterns}
\end{table}

\section{Discussion}
The results extend prior work showing that LLMs can produce useful research-quality scores under some conditions \citep{thelwall2024chatgpt-quality,thelwall2025settings-inputs,thelwall-yang2025probabilities,thelwall-mohammadi2026small-reasoning}. The main contribution is to show that score guidance does not merely make the model copy a supplied score. Instead, the supplied score can act as an anchor that influences both the final score and the review rationale in corrected cases.

This rationale-steering effect was especially visible for baseline underestimates. When guidance raised these scores successfully, the review frame also shifted toward a more positive evaluation. Downward correction was less straightforward: overestimates were more likely to persist when the rationale retained strong positive framing or failed to incorporate downgrade evidence used by human reviewers. One possible explanation is that upward correction may align with the positively framed style often produced by instruction-tuned LLMs. Downward correction, by contrast, requires the model to incorporate more specific downgrade evidence, which was less consistently covered in the point-level analysis. This refines prior work on overlap between LLM and human scientific feedback~\citep{liang2024can}: the important issue is not only whether similar topics are mentioned, but whether they are used in the same scoring direction.

Overall, score guidance should be understood as a useful but incomplete rationale-steering mechanism. It can improve some generated reviews as well as scores, especially when the required change is consistent with a more positive evaluation frame. However, expert downgrade evidence remains harder to incorporate reliably. Future score-guided review systems should therefore be evaluated by whether the review rationale changes in the direction required for the target score, not only by whether the final score is correct~\citep{thelwall-kurt2025bias,thelwall2026responsible}.

\section*{Limitations}
This study has several limitations. First, the dataset is small, domain-specific, and shaped by self-selection. The 98 anonymised health-domain outputs came from an internal REF-style evaluation process, meaning that they were submitted or selected for assessment as potential high research-quality candidates rather than randomly sampled from all publications. As a result, the scores are concentrated around the 3* level, with relatively few very low-quality outputs. The results should therefore be interpreted as evidence from a controlled case study of REF-candidate-like outputs, rather than as general estimates for all research evaluation settings or the full 1--4 score range. Second, the guided condition uses oracle score guidance, where the supplied score is set to the rounded human reference score. This is useful for testing whether a correct signal can affect both scores and rationales, but it does not capture the additional uncertainty of using predicted scores in a deployed system. Third, the text-level analysis relies on automatic annotation, including zero-shot sentence-level labels and Qwen-based evaluation-point extraction and coverage judgements. These annotations enable scalable comparison of review rationales, but future work should validate them against expert-coded evidence points. 


\bibliography{custom}

\clearpage
\onecolumn
\appendix
\renewcommand{\thefigure}{A\arabic{figure}}
\setcounter{figure}{0}
\renewcommand{\thetable}{A\arabic{table}}
\setcounter{table}{0}
\section{Appendix}
\label{sec:appendix}
\subsection{Reproducibility Details}
\label{app:reproducibility}

\subsubsection{Dataset and preprocessing}\label{appendix:dataset}
The dataset was derived from an internal REF-style research evaluation process at the University of XXX. The data are difficult to obtain because they include private assessment reports and scores provided by specialist internal reviewers and academic staff. The outputs were not a random sample of all publications. They were submitted for internal REF-style assessment, and therefore mainly represent research outputs considered plausible candidates for evaluation rather than the full quality range of published work. 

Each submitted output was assessed by two to four reviewers, who provided qualitative comments and recommended quality scores. The intended basis for assessment was the content of the output itself, rather than author, institution, journal, or venue information. These reviews were then used within the internal assessment process to produce an adjudicated reference score on a 1--4 scale. The reports and scores were private internal evaluation materials and were not publicly available.

For this study, 98 anonymised health-domain research outputs with associated human review reports and adjudicated reference scores were used. The submitted articles were downloaded as PDF files. Before text extraction, author-identifying and publication-related information was manually removed where possible to reduce the chance that LLMs could rely on author, institution, or venue signals rather than paper content.

Full-text content was extracted from the cleaned PDFs using PDFMiner, a Python-based PDF parsing and text-extraction tool. Each extracted file was manually checked and cleaned to remove junk text from complex formulae, images, and tables. Section information was extracted with GROBID~\citep{lopez2009grobid} and post-processed with custom Python scripts, which merged subsections into top-level sections using textual cues such as section numbering, section headings, and capitalisation patterns. These top-level sections were used to construct the LLM input variants: title and abstract; title, introduction, and conclusion; and full text. The main text-level analyses use the full-text condition, while the broader output set also included the two shorter input conditions. All outputs and review reports were in English.

\subsubsection{Prompt templates}\label{appendix:prompt_templates}
The generation prompt consisted of a shared system prompt and a task prompt. The same system prompt was used in all conditions (Figure~\ref{fig:system_prompt}). The no-guidance task prompt was used for baseline review generation (Figure~\ref{fig:no_guidance_prompt}), while the oracle-guidance task prompt included an external score block (Figure~\ref{fig:guided_prompt}). Three prompt variants were used, differing only in the order of the three prompt blocks: v4 = rule--format--paper text, v7 = paper text--rules--format, and v9 = rules--paper text--format. The evaluation criteria and required output fields were kept the same.

\begin{figure}[t]
\centering
\begin{tcblisting}{
  listing only,
  width=0.95\textwidth,
  colback=gray!5,
  colframe=black!75,
  colbacktitle=black!75,
  coltitle=white,
  title=Shared system prompt,
  fonttitle=\bfseries\small,
  arc=2mm,
  boxrule=0.6pt,
  left=2mm,
  right=2mm,
  top=1mm,
  bottom=1mm,
  listing options={
    basicstyle=\ttfamily\tiny,
    breaklines=true,
    breakatwhitespace=true,
    breakindent=0pt,
    breakautoindent=false,
    columns=fullflexible,
    keepspaces=true,
    showstringspaces=false
  }
}
You are an academic expert, assessing academic journal articles based on originality, significance, and rigour in alignment with international research quality standards. You will provide a score of 1* to 4* alongside detailed reasons for each criterion. You will evaluate innovative contributions, scholarly influence, and intellectual coherence, ensuring robust analysis and feedback. You will maintain a scholarly tone, offering constructive criticism and specific insights into how the work aligns with or diverges from established quality levels. You will emphasize scientific rigour, contribution to knowledge, and applicability in various sectors, providing comprehensive evaluations and detailed explanations for its scoring.

**Originality** will be understood as the extent to which the output makes an important and innovative contribution to understanding and knowledge in the field. Research outputs that demonstrate originality may do one or more of the following: produce and interpret new empirical findings or new material; engage with new and/or complex problems; develop innovative research methods, methodologies and analytical techniques; show imaginative and creative scope; provide new arguments and/or new forms of expression, formal innovations, interpretations and/or insights; collect and engage with novel types of data; and/or advance theory or the analysis of doctrine, policy or practice, and new forms of expression.
**Significance** will be understood as the extent to which the work has influenced, or has the capacity to influence, knowledge and scholarly thought, or the development and understanding of policy and/or practice. 
**Rigour** will be understood as the extent to which the work demonstrates intellectual coherence and integrity, and adopts robust and appropriate concepts, analyses, sources, theories and/or methodologies.

The scoring system used is 1*, 2*, 3* or 4*, which are defined as follows.
- 4*: Quality that is world-leading in terms of originality, significance and rigour.
- 3*: Quality that is internationally excellent in terms of originality, significance and rigour but which falls short of the highest standards of excellence.
- 2*: Quality that is recognised internationally in terms of originality, significance and rigour.
- 1*: Quality that is recognised nationally in terms of originality, significance and rigour.

The terms 'world-leading', 'international' and 'national' will be taken as quality benchmarks within the generic definitions of the quality levels. They will relate to the actual, likely or deserved influence of the work, whether in the UK, a particular country or region outside the UK, or on international audiences more broadly. There will be no assumption of any necessary international exposure in terms of publication or reception, or any necessary research content in terms of topic or approach. Nor will there be an assumption that work published in a language other than English or Welsh is necessarily of a quality that is or is not internationally benchmarked. 

In assessing outputs, look for evidence of originality, significance and rigour and apply the generic definitions of the starred quality levels as follows:
- In assessing work as being 4* (quality that is world-leading in terms of originality, significance and rigour), expect to see evidence of, or potential for, some of the following types of characteristics across and possibly beyond its area/field: a primary or essential point of reference; of profound influence; instrumental in developing new thinking, practices, paradigms, policies or audiences; a major expansion of the range and the depth of research and its application; outstandingly novel, innovative and/or creative.
- In assessing work as being 3* (quality that is internationally excellent in terms of originality, significance and rigour but which falls short of the highest standards of excellence), expect to see evidence of, or potential for, some of the following types of characteristics across and possibly beyond its area/field: an important point of reference; of considerable influence; a catalyst for, or important contribution to, new thinking, practices, paradigms, policies or audiences; a significant expansion of the range and the depth of research and its application; significantly novel or innovative or creative.
- In assessing work as being 2* (quality that is recognised internationally in terms of originality, significance and rigour), expect to see evidence of, or potential for, some of the following types of characteristics across and possibly beyond its area/field: a recognised point of reference; of some influence; an incremental and cumulative advance on thinking, practices, paradigms, policies or audiences; a useful contribution to the range or depth of research and its application.
- In assessing work as being 1* (quality that is recognised nationally in terms of originality, significance and rigour), expect to see evidence of the following characteristics within its area/field: an identifiable contribution to understanding without advancing existing paradigms of enquiry or practice; of minor influence.
\end{tcblisting}
\caption{Shared system prompt used for all LLM review generation.}
\label{fig:system_prompt}
\end{figure}

\begin{figure*}[t]
\centering
\begin{tcblisting}{
  listing only,
  width=0.95\textwidth,
  colback=gray!5,
  colframe=black!75,
  colbacktitle=black!75,
  coltitle=white,
  title=No guidance task prompt,
  fonttitle=\bfseries\small,
  arc=2mm,
  boxrule=0.6pt,
  left=2mm,
  right=2mm,
  top=1mm,
  bottom=1mm,
  listing options={
    basicstyle=\ttfamily\tiny,
    breaklines=true,
    breakatwhitespace=true,
    breakindent=0pt,
    breakautoindent=false,
    columns=fullflexible,
    keepspaces=true,
    showstringspaces=false
  }
}
[RULES]
Please provide a formal academic peer review of the provided paper.
### Instructions:
1. You MUST evaluate the paper across three dimensions: Originality, Significance, and Rigour. For EACH dimension, provide a detailed qualitative report followed by a quantitative score (1 to 4).
2. The qualitative report for each section MUST be written as a single, cohesive paragraph that is academically reasoned and grounded in specific evidence from the paper.
3. You MUST provide an ``Overall Score'' as a single integer from 1 to 4. This score must logically reflect the evaluation provided in the preceding sections.
4. You MUST NOT stop early or omit any section.
5. You MUST NOT include any content outside the specified output format.
[/RULES]

[OUTPUT_FORMAT(Exactly, including order, headings, and punctuation)]
Originality report:
Originality score:
Significance report:
Significance score:
Rigour report:
Rigour score:
Overall report:
Overall score:
[/OUTPUT_FORMAT]

[PAPER]
{{PAPER_TEXT}}
[/PAPER]
\end{tcblisting}
\caption{No-guidance task prompt.}
\label{fig:no_guidance_prompt}
\end{figure*}

\begin{figure}[t]
\centering
\begin{tcblisting}{
  listing only,
  width=0.95\textwidth,
  colback=gray!5,
  colframe=black!75,
  colbacktitle=black!75,
  coltitle=white,
  title=Guided task prompt,
  fonttitle=\bfseries\small,
  arc=2mm,
  boxrule=0.6pt,
  left=2mm,
  right=2mm,
  top=1mm,
  bottom=1mm,
  listing options={
    basicstyle=\ttfamily\tiny,
    breaklines=true,
    breakatwhitespace=true,
    breakindent=0pt,
    breakautoindent=false,
    columns=fullflexible,
    keepspaces=true,
    showstringspaces=false
  }
}
[RULES]
Please provide a formal academic peer review of the provided paper.

### External Anchor
[SCORE_BLOCK]
A specialized SciBERT model has pre-screened this paper and assigned a score of: {{GUIDE_SCORE}}.
Note: Consider the SciBERT score as the Bayesian prior for this evaluation. Adjust this prior only upon encountering substantial evidence within the text.
[/SCORE_BLOCK]

### Instructions:
1. You MUST evaluate the paper across three dimensions: Originality, Significance, and Rigour. For EACH dimension, provide a detailed qualitative report followed by a quantitative score (1 to 4).
2. The qualitative report for each section MUST be written as a single, cohesive paragraph that is academically reasoned and grounded in specific evidence from the paper.
3. You MUST provide an ``Overall Score'' as a single integer from 1 to 4. This score must logically reflect the evaluation provided in the preceding sections.
4. You MUST NOT stop early or omit any section.
5. You MUST NOT include any content outside the specified output format.
[/RULES]

[OUTPUT_FORMAT(Exactly, including order, headings, and punctuation)]
Originality report:
Originality score:
Significance report:
Significance score:
Rigour report:
Rigour score:
Overall report:
Overall score:
[/OUTPUT_FORMAT]

[PAPER]
{{PAPER_TEXT}}
[/PAPER]
\end{tcblisting}
\caption{Guided task prompt used for score-guided review generation. The placeholder \texttt{\{\{GUIDE\_SCORE\}\}} was filled with guide scores from 1 to 4; the oracle-guided subset used the rounded human reference score.}
\label{fig:guided_prompt}
\end{figure}

\subsubsection{Model, decoding, and configuration screening}\label{appendix:screening}
Outputs were generated with three model settings: Llama-3.1-8B-Instruct, Falcon3-10B-Instruct, and Llama-3.1-405B-Instruct. For Llama-3.1-8B and Falcon3-10B, three prompt variants were used (v4, v7, and v9). For Llama-3.1-405B, only prompt v4 was used because of generation cost.

The broader output set included three input-length conditions: title and abstract; title, introduction, and conclusion; and full text. For each input condition, outputs were generated under 11 decoding settings: greedy decoding, and sampling with temperature values of 0.1, 0.3, 0.5, 0.7, and 0.9 crossed with top-p values of 0.8 and 0.9. This produced 33 input--decoding configurations for each model/prompt setting. Configuration screening was then applied to this full candidate set.

To reduce redundancy across generation configurations, no-guidance baseline outputs were screened for shared error patterns, following the idea that multiple LLM outputs may contain correlated rather than independent error information~\citep{kohli2026nine}. For each candidate configuration, a signed score-difference vector was constructed by subtracting the rounded human reference score from the LLM score for each paper. Configurations were then compared using normalized mutual information between these signed error vectors. Configurations with lower shared error information and acceptable score quality were preferred.

The main text analysis used three selected full-text configurations. Full-text input was retained because it provided the richest paper evidence and was consistently among the stronger candidates in the screening. The same three decoding settings were retained across model and prompt settings: temperature/top-p = 0.1/0.9, 0.3/0.8, and 0.5/0.9. The main paper reports Llama-3.1-8B prompt v4 as the primary text-analysis setting, while the corresponding results for the other model and prompt settings are used as robustness checks. Table~\ref{tab:app_selected_configs} reports the selected configurations and their screening statistics.

\begin{table}[h]
\centering
\small
\setlength{\tabcolsep}{4pt}
\begin{tabular}{llrrrr}
\hline
Input & Decoding setting & Acc. & MAE & Avg. NMI & Info. contrib. \\
\hline
Full text & temp=0.5, top-p=0.9 & 0.469 & 0.571 & 0.146 & 0.106 \\
Full text & temp=0.1, top-p=0.9 & 0.582 & 0.449 & 0.155 & 0.089 \\
Full text & temp=0.3, top-p=0.8 & 0.418 & 0.622 & 0.175 & 0.047 \\
\hline
\end{tabular}
\caption{Selected decoding configurations for the main baseline--guided text analysis. Avg. NMI is the average normalized mutual information between a configuration's signed score-difference vector and the vectors from other candidate configurations.}
\label{tab:app_selected_configs}
\end{table}

\subsubsection{Annotation labels, prompts, and text features}\label{appendix:text_features}
Text-derived features were constructed from three annotation layers: sentence-level zero-shot review-frame labels, Qwen-extracted evaluation points, and Qwen coverage judgements for human evaluation points.

\paragraph{Sentence-level zero-shot labels.}
Sentence-level labels were assigned with two zero-shot classification tasks. The stance task used the hypothesis template: \texttt{This review sentence expresses a \{\}} with the labels: strength of the paper; weakness or limitation of the paper; mixed evaluation of the paper; neutral description of the paper. The review-frame theme task used the hypothesis template: \texttt{This review sentence discusses \{\}} with the labels: international or global significance; novel or world-leading contribution; robust methodology or strong evidence; practical clinical industrial or policy impact; limited sample size or preliminary evidence; limited generalisability or narrow context; methodological weakness or missing detail; incremental or uncertain contribution.

\paragraph{Qwen point-level annotation.}
Qwen3-32B was used for point extraction and coverage judgement. First, it extracted paper-specific evaluation points from human review reports using the prompt shown in Figure~\ref{fig:qwen_human_point_extraction_prompt}. Each point was labelled by aspect, theme, stance, and score implication. The point-level theme labels were developed after manually inspecting a subset of human reviews and were not identical to the three broad REF criteria. They were: novelty or originality; international or global significance; clinical practical policy or industrial impact; methodology or rigour; evidence scale or dataset strength; generalisability or context limitation; incremental or uncertain contribution; clarity or presentation; other. LLM review points were extracted with the same point schema, replacing the human review input with the LLM-generated review report. Second, the extracted human evaluation points were used as the reference for coverage checking. Qwen judged whether each human point was covered, partially covered, or not covered in the corresponding LLM review, using the prompt shown in Figure~\ref{fig:qwen_coverage_prompt}. The coverage prompt did not assign theme or score-implication labels; these labels came from the preceding point-extraction step.

\begin{figure*}[t]
\centering
\begin{tcblisting}{
  listing only,
  width=0.95\textwidth,
  colback=gray!5,
  colframe=black!75,
  colbacktitle=black!75,
  coltitle=white,
  title=Qwen human evaluation-point extraction prompt,
  fonttitle=\bfseries\small,
  arc=2mm,
  boxrule=0.6pt,
  left=2mm,
  right=2mm,
  top=1mm,
  bottom=1mm,
  listing options={
    basicstyle=\ttfamily\tiny,
    breaklines=true,
    breakatwhitespace=true,
    breakindent=0pt,
    breakautoindent=false,
    columns=fullflexible,
    keepspaces=true,
    showstringspaces=false
  }
}
System prompt:
You are a careful research-quality review analyst. You extract and compare evaluation points for originality, significance, and rigour. Return valid JSON only. Do not include reasoning, markdown, or commentary.

User prompt:
Task: Extract paper-specific evaluation points from the human expert reviews.

Only use information explicitly stated in the human reviews. Do not infer from the paper title or from outside knowledge.
Merge duplicate points across reviewers. Prefer concrete points about originality, significance, rigour, limitations, evidence, methods, impact, or world-leading contribution.
Avoid generic points such as "good paper" unless the review gives a specific reason.
Do not extract points that are only about the score label, REF eligibility/authorship, reviewer confidence/expertise, or whether the output meets formal criteria, unless the same sentence also gives a concrete research-quality reason.
Extract at most {{MAX_POINTS}} points.

Labels:
- aspect: originality, significance, rigour, overall_or_other
- theme: choose one short label from this list: {{THEME_LABELS}}
- stance: strength, weakness, mixed, neutral
- importance: high, medium, low
- score_implication: higher_score, lower_score, mixed_or_unclear, neutral

Return JSON only, with this exact structure:
{
  "paper_id": "string",
  "points": [
    {
      "point_id": "P1",
      "point_text": "one concise paper-specific evaluation point",
      "aspect": "originality|significance|rigour|overall_or_other",
      "theme": "short_theme_label",
      "stance": "strength|weakness|mixed|neutral",
      "importance": "high|medium|low",
      "score_implication": "higher_score|lower_score|mixed_or_unclear|neutral",
      "key_phrases": ["short phrase from the human review"]
    }
  ]
}

paper_id: {{PAPER_ID}}

Human reviews:
{{HUMAN_REVIEW_TEXT}}
\end{tcblisting}
\caption{Qwen prompt used to extract evaluation points from human review reports.}
\label{fig:qwen_human_point_extraction_prompt}
\end{figure*}

\begin{figure*}[t]
\centering
\begin{tcblisting}{
  listing only,
  width=0.95\textwidth,
  colback=gray!5,
  colframe=black!75,
  colbacktitle=black!75,
  coltitle=white,
  title=Qwen human-point coverage judgement prompt,
  fonttitle=\bfseries\small,
  arc=2mm,
  boxrule=0.6pt,
  left=2mm,
  right=2mm,
  top=1mm,
  bottom=1mm,
  listing options={
    basicstyle=\ttfamily\tiny,
    breaklines=true,
    breakatwhitespace=true,
    breakindent=0pt,
    breakautoindent=false,
    columns=fullflexible,
    keepspaces=true,
    showstringspaces=false
  }
}
System prompt:
You are a careful research-quality review analyst. You extract and compare evaluation points for originality, significance, and rigour. Return valid JSON only. Do not include reasoning, markdown, or commentary.

User prompt:
Task: For each human evaluation point, check whether each LLM report covers that point.

Important rules:
- Compare only the human point with the LLM report for the same paper.
- "covered" means the LLM report clearly expresses the same evaluation point with enough detail.
- "partially_covered" means the LLM report mentions the same broad idea but is generic, weaker, or misses an important detail.
- "not_covered" means the LLM report does not mention the point.
- If the LLM mentions the point but evaluates it differently, use coverage_status as covered or partially_covered, and set stance_relation to opposite, weaker, or stronger as appropriate.
- Use short evidence_quote from the LLM report only. If not covered, evidence_quote should be "".
- Use coverage_score 1.0 for covered, 0.5 for partially_covered, and 0.0 for not_covered.
- Return one coverage row for every report_id and point_id combination.

Allowed stance_relation values:
same, weaker, stronger, opposite, missing, unclear

Allowed reason_code values:
covered, generic_only, partial_detail, missing_point, weaker_strength, stronger_strength, opposite_stance, unclear

Return JSON only, with this exact structure:
{
  "paper_id": "string",
  "coverage": [
    {
      "report_id": "string",
      "point_id": "P1",
      "coverage_status": "covered|partially_covered|not_covered",
      "stance_relation": "same|weaker|stronger|opposite|missing|unclear",
      "llm_stance": "strength|weakness|mixed|neutral|missing|unclear",
      "coverage_score": 1.0,
      "reason_code": "covered|generic_only|partial_detail|missing_point|weaker_strength|stronger_strength|opposite_stance|unclear",
      "evidence_quote": "short quote from the LLM report, or empty string"
    }
  ]
}

paper_id: {{PAPER_ID}}

Human evaluation points:
{{HUMAN_EVALUATION_POINTS_JSON}}

LLM reports to check:
{{LLM_REPORTS_JSON}}
\end{tcblisting}
\caption{Qwen prompt used to judge whether human evaluation points were covered in the corresponding LLM review.}
\label{fig:qwen_coverage_prompt}
\end{figure*}

\paragraph{Text-derived features.}
Table~\ref{tab:app_text_features} summarises the main text-derived feature groups used in the Results section. Sentence-level review-frame features are calculated from LLM review sentences, while Qwen-based features are calculated over extracted evaluation points. For baseline--guided comparisons, delta features are calculated as guided minus baseline. Theme distributions are computed within each source and score-implication group, as in Table~\ref{tab:evidence_frames}. Human-point coverage uses human evaluation points as the reference. Stance and score implication are treated separately: stance captures evaluative direction, while score implication captures whether the point supports a higher or lower score.

\begin{table}[t]
\centering
\small
\begin{tabular}{p{0.18\linewidth}p{0.35\linewidth}p{0.20\linewidth}p{0.15\linewidth}}
\hline
Feature group & Feature examples & Calculation & Method/model \\
\hline
Review-frame stance &
Strength ratio; weakness ratio; strength--weakness balance &
$n_{\mathrm{sent}}(\mathrm{label}) / n_{\mathrm{sent}}$ &
Zero-shot classifier \\

Review-frame theme &
Novelty/world-leading ratio; methodological-weakness ratio; generalisability-limitation ratio; practical-impact ratio; etc. &
$n_{\mathrm{sent}}(\mathrm{theme}) / n_{\mathrm{sent}}$ &
Zero-shot classifier \\

Evaluation-point theme distribution &
Novelty/originality; practical impact; international significance; methodology/rigour; generalisability limitation; incremental/uncertain contribution; evidence scale; other &
$n_{\mathrm{point}}(\mathrm{theme}) / n_{\mathrm{point}}$, within each source and score-implication group &
Qwen3-32B \\

Human-point coverage &
Coverage of human strength points; coverage of human weakness points; theme-specific coverage &
$\frac{1}{N}\sum_{i=1}^{N} c_i$, where $c_i \in \{1, 0.5, 0\}$ &
Qwen3-32B \\
\hline
\end{tabular}
\caption{Main text-derived evidence and coverage features used in the Results section. Sentence-level features are calculated over review sentences; point-level features are calculated over extracted evaluation points.}
\label{tab:app_text_features}
\end{table}

\subsubsection{Compute resources and runtime}
The pipeline had three computational stages: LLM review generation, sentence-level zero-shot labelling, and Qwen-based point-level annotation. All jobs were submitted as Slurm batch jobs on the university HPC cluster. One generation output file corresponds to one model--prompt--input--decoding--guidance setting run over the 98 papers.

\begin{table}[h]
\centering
\small
\setlength{\tabcolsep}{4pt}
\begin{tabular}{p{0.16\textwidth} p{0.32\textwidth} p{0.24\textwidth} p{0.22\textwidth}}
\hline
Stage & Model / method & Scope & Approximate runtime \\
\hline
Review generation &
\texttt{meta-llama/Llama-3.1-8B-Instruct}; \texttt{tiiuae/Falcon3-10B-Instruct} &
For each model/prompt/input setting: 11 no-guidance files and 44 guided files &
About 20 minutes per output file; about 18.3 hours per model/prompt/input setting if run sequentially \\

 &
Llama-3.1-405B via Ollama &
Prompt v4 with three input lengths; 11 no-guidance and 44 guided files per input length &
About 2.5 hours per output file using four NVIDIA H100 GPUs \\

Sentence-level labelling &
DeBERTa-v3-large zero-shot classifier &
Stance and review-frame labels for one model/prompt version &
Within approximately one hour \\

Point-level annotation &
Qwen3-32B as an LLM judge &
Evaluation-point extraction and human-point coverage checking for one model/prompt version &
Approximately five hours \\
\hline
\end{tabular}
\caption{Approximate compute requirements for generation and annotation. Runtimes are wall-clock estimates and varied with queue conditions and input length.}
\label{tab:app_compute_runtime}
\end{table}

For the 8B and 10B models, each model/prompt/input setting produced 55 output files: 11 no-guidance files and 44 guided files, corresponding to four supplied guide scores for each decoding setting. At approximately 20 minutes per file, this is about 18.3 sequential hours per model/prompt/input setting. With three prompt variants and three input lengths, this corresponds to 495 output files for Llama-3.1-8B and 495 output files for Falcon3-10B.

The Llama-3.1-405B setting used prompt v4 and three input lengths, giving 165 output files in total. At approximately 2.5 hours per file, this corresponds to about 412.5 sequential GPU-job hours. Actual wall-clock completion time was shorter when jobs were submitted in parallel, but depended on HPC scheduling.

The annotation stages were run after review generation. Full Slurm scripts, runtime logs, prompt files, and model settings are retained with the supplementary material.

\subsection{Supplementary results}
\label{app:supp_results}
\subsubsection{Supplementary score-guidance results}
Table~\ref{tab:app_full_score_guidance} reports the broader score-level results underlying the selected full-text results in the main paper. Unlike Table~\ref{tab:score_guidance_accuracy}, which focuses on the selected configurations used for text analysis, this table summarises the wider generated output set across input lengths and decoding settings. For each setting, baseline accuracy and MAE are computed from no-guidance outputs, while guided accuracy and MAE are computed from the oracle-guided subset where the supplied guide score equals the rounded human reference score.

\begin{table}[t]
\centering
\small
\setlength{\tabcolsep}{3pt}
\begin{tabular}{llrrrrrrr}
\hline
Model & Prompt & $n$ & Acc. base & Acc. guid. & $\Delta$Acc. & MAE base & MAE guid. & $\Delta$MAE \\
\hline
Llama-8B & v4 & 3234 & 0.438 & 0.483 & +0.045 & 0.629 & 0.560 & -0.070 \\
Llama-8B & v7 & 3233 & 0.414 & 0.594 & +0.180 & 0.657 & 0.437 & -0.220 \\
Llama-8B & v9 & 3234 & 0.420 & 0.486 & +0.066 & 0.642 & 0.554 & -0.089 \\
Falcon-10B & v4 & 3234 & 0.638 & 0.793 & +0.154 & 0.367 & 0.209 & -0.158 \\
Falcon-10B & v7 & 3234 & 0.622 & 0.777 & +0.155 & 0.386 & 0.224 & -0.162 \\
Falcon-10B & v9 & 3234 & 0.657 & 0.816 & +0.159 & 0.344 & 0.185 & -0.159 \\
Llama-405B & v4 & 3230 & 0.595 & 0.826 & +0.231 & 0.429 & 0.175 & -0.254 \\
\hline
\end{tabular}
\caption{Supplementary oracle-guidance score results across the broader output set. Results are aggregated across three input lengths and 11 decoding settings. Guided scores are computed from the oracle-guided subset where the supplied guide score equals the rounded human reference score. Negative $\Delta$MAE values indicate reduced distance from the human reference score after guidance.}
\label{tab:app_full_score_guidance}
\end{table}

Table~\ref{tab:app_score_following} reports score-following checks using the full guide-score sweep from 1 to 4. These checks test whether the final score was fully determined by the supplied guide score.

\begin{table}[t]
\centering
\small
\setlength{\tabcolsep}{4pt}
\begin{tabular}{lrrrr}
\hline
Setting & Guide-copy rate & Corr. guide--final & All-four copied & $n$ guided cases \\
\hline
Main setting & 0.319 & 0.322 & 0.000 & 1176 \\
Broader score set & 0.359 & 0.374 & 0.001 & 12936 \\
\hline
\end{tabular}
\caption{Score-following checks using supplied guide scores from 1 to 4. The main setting is Llama-8B v4 with full-text input and three selected decoding configurations. The broader score set uses the same model and prompt, but includes three input lengths and 11 decoding settings. Guide-copy rate is the proportion of guided cases where the final LLM score exactly matched the supplied guide score. Corr.~guide--final is the correlation between the supplied guide score and the final LLM score. All-four copied indicates the proportion of paper--configuration cases for which all four supplied guide scores were copied exactly.}
\label{tab:app_score_following}
\end{table}

\subsubsection{Score transition results}
Table~\ref{tab:app_score_transitions} reports score transitions for each selected model, prompt version, and decoding setting. For each row, baseline and oracle-guided scores were classified as underestimates, correct matches, or overestimates relative to the rounded human reference score. The table reports the number of baseline mismatches, the number corrected after guidance, the number that remained mismatched in the same direction, and the number of initially correct cases that became mismatched after guidance.

\begin{table*}[t]
\centering
\small
\setlength{\tabcolsep}{3pt}
\begin{tabular}{lllrrrrr}
\hline
Model & Pr. & Decoding & Baseline wrong & Corrected & Persistent & New err. & Acc. guid. \\
\hline
Llama-8B & v4 & 0.1/0.9 & 41 & 21 & 19 & 21 & 0.582 \\
Llama-8B & v4 & 0.3/0.8 & 56 & 23 & 30 & 12 & 0.541 \\
Llama-8B & v4 & 0.5/0.9 & 53 & 25 & 23 & 18 & 0.531 \\
Llama-8B & v7 & 0.1/0.9 & 42 & 27 & 15 & 10 & 0.745 \\
Llama-8B & v7 & 0.3/0.8 & 42 & 27 & 15 & 9 & 0.755 \\
Llama-8B & v7 & 0.5/0.9 & 42 & 31 & 10 & 7 & 0.816 \\
Llama-8B & v9 & 0.1/0.9 & 62 & 31 & 28 & 14 & 0.541 \\
Llama-8B & v9 & 0.3/0.8 & 48 & 20 & 25 & 15 & 0.561 \\
Llama-8B & v9 & 0.5/0.9 & 51 & 24 & 26 & 16 & 0.561 \\
Falcon-10B & v4 & 0.1/0.9 & 47 & 24 & 23 & 2 & 0.745 \\
Falcon-10B & v4 & 0.3/0.8 & 51 & 26 & 25 & 3 & 0.714 \\
Falcon-10B & v4 & 0.5/0.9 & 40 & 20 & 20 & 9 & 0.704 \\
Falcon-10B & v7 & 0.1/0.9 & 49 & 21 & 27 & 3 & 0.684 \\
Falcon-10B & v7 & 0.3/0.8 & 52 & 23 & 29 & 5 & 0.653 \\
Falcon-10B & v7 & 0.5/0.9 & 54 & 28 & 26 & 7 & 0.663 \\
Falcon-10B & v9 & 0.1/0.9 & 35 & 16 & 19 & 0 & 0.806 \\
Falcon-10B & v9 & 0.3/0.8 & 38 & 20 & 18 & 2 & 0.796 \\
Falcon-10B & v9 & 0.5/0.9 & 38 & 22 & 16 & 5 & 0.786 \\
Llama-405B & v4 & 0.1/0.9 & 37 & 27 & 10 & 3 & 0.867 \\
Llama-405B & v4 & 0.3/0.8 & 33 & 21 & 12 & 3 & 0.847 \\
Llama-405B & v4 & 0.5/0.9 & 36 & 28 & 8 & 10 & 0.816 \\
\hline
\end{tabular}
\caption{Score transition summaries by model, prompt version, and selected decoding setting. Each row summarises 98 papers under one model--prompt--decoding setting. Baseline wrong is the number of baseline scores that differed from the rounded human reference score. Corrected is the number of baseline mismatches corrected by oracle guidance. Persistent is the number of baseline mismatches that remained mismatched in the same direction. New err. denotes cases that matched the human reference score at baseline but became mismatched after oracle guidance.}
\label{tab:app_score_transitions}
\end{table*}

\subsubsection{Summary of robustness checks across settings}
The main text-level analysis used Llama-8B with prompt v4 as the primary setting. Robustness checks were then conducted with the other selected model and prompt settings: Llama-8B v7/v9, Falcon-10B v4/v7/v9, and Llama-405B v4. These checks assess whether the direction of the main findings is stable across settings, rather than whether every model produces identical numerical effects.

Table~\ref{tab:app_robustness_summary} summarises these checks. Score-level effects were the most stable. Text-level effects were also broadly consistent, especially the link between LLM scores and review frames, but the coverage of specific human downgrade evidence was more model-dependent.

\begin{table*}[t]
\centering
\small
\setlength{\tabcolsep}{4pt}
\begin{tabular}{p{0.29\textwidth} p{0.16\textwidth} p{0.30\textwidth} p{0.18\textwidth}}
\hline
Finding checked & Support & Key evidence & Interpretation \\
\hline
Oracle guidance improved scoring &
7/7 settings &
Accuracy increased and MAE decreased for all selected model/prompt settings. &
Stable score-level effect \\

Guided scores were not fully copied &
7/7 settings &
Guide-copy rate = 31.89--48.38\%; all-four-copy rate = 0--0.34\%. &
Score-sensitive, not deterministic copying \\

Baseline mismatches were partly corrected but often persisted &
21/21 rows &
505 baseline mismatches corrected; 424 persisted in the same direction. &
Persistent mismatches were not specific to the main setting \\

LLM scores aligned with their own review frame &
5/6 additional settings &
Baseline scores generally rose with strength framing and fell with weakness or limitation framing. &
Mostly stable text-level pattern \\

Correction was associated with review-frame change &
5/6 additional settings &
Corrected mismatches usually showed larger frame shifts in the expected direction than persistent mismatches. &
Mostly stable, feature contrasts varied \\

Human strength points were covered more reliably than weakness points &
Cross-setting aggregate &
Strength/upgrade coverage = 0.761--0.787; downgrade/weakness coverage = 0.204--0.240. &
Stable point-level asymmetry \\

Persistent overestimates missed downgrade evidence &
Mixed support &
Strongest in Llama-8B v4; present in several but not all additional settings. &
Suggestive and model-dependent \\
\hline
\end{tabular}
\caption{Summary of robustness checks across selected model and prompt settings. The additional settings are Llama-8B v7/v9, Falcon-10B v4/v7/v9, and Llama-405B v4.}
\label{tab:app_robustness_summary}
\end{table*}

\end{document}